\documentclass[10pt]{iopart}
\newcommand{\gguide}

\usepackage{multicol}                  
\usepackage{epsfig}                                                            
\usepackage{graphicx}
\usepackage{dcolumn}
\usepackage{color}
\usepackage{bm}
\usepackage{subfigure}

\begin{document}

\title{\bf Charge-transfer-driven enhanced room-temperature ferromagnetism in BiFeO$_3$/Ag nanocomposite }

\author {Tania Chatterjee,$^{1,2}$ Shubhankar Mishra,$^3$ Arnab Mukherjee,$^2$ Prabir Pal,$^4$ Biswarup Satpati$^5$ and Dipten Bhattacharya$^1$} \address {$^1$Advanced Materials and Chemical Characterization Division, CSIR-Central Glass and Ceramic Research Institute, Kolkata 700032, India} \address {$^2$Functional Materials and Device Division, CSIR-Central Glass and Ceramic Research Institute, Kolkata 700032, India}
\address {$^3$School of Materials Science and Nanotechnology, Jadavpur University, Kolkata 700032, India; Current address: Department of Physics, Indian Institute of Science, Bangalore 560012, India} 
\address {$^4$Material Characterization Division, CSIR-Central Glass and Ceramic Research Institute, Kolkata 700032, India}
\address {$^5$Saha Institute of Nuclear Physics, a CI of Homi Bhabha National Institute, 1/AF Salt Lake, Kolkata 700064, India}

\ead{arnabm@cgcri.res.in}

\date{\today}

\begin{abstract}
We report observation of more than an order of magnitude jump in saturation magnetization in BiFeO$_3$/Ag nanocomposite at room temperature compared to what is observed in bare BiFeO$_3$ nanoparticles. Using transmission electron microscopy together with energy dispersive x-ray spectra (which maps the element concentration across the BiFeO$_3$/Ag interface) and x-ray photoelectron spectroscopy, we show that both the observed specific self-assembly pattern of BiFeO$_3$ and Ag nanoparticles and the charge transfer between Ag and O are responsible for such an enormous rise in room-temperature magnetization. The BiFeO$_3$/Ag nanocomposites, therefore, could prove to be extremely useful for a variety of applications including biomedical.\\


\end{abstract}

\maketitle

\section{Introduction}
Decoration of malignant tumors with magnetic nanoparticles results in enhanced efficacy of the treatment by hyperthermia \cite{Hoopes}. In this context, large room-temperature magnetization and soft ferromagnetism in a variety of bare or surface-functionalized oxide or alloy magnetic nanoparticles turns out to be especially useful. In recent time, nanoscale Fe$_{65}$Co$_{35}$ or Fe$_{100-x}$Co$_x$ intermetallic alloys \cite{Vera-Serna} (saturation magnetization $M_S$ $\sim$240 emu/g) or iron oxide nanoparticles \cite{Woo,Laurent} ($M_S$ $\sim$70-80 emu/g) were used for such purposes. However, alloy systems are difficult to synthesize and often oxidize. Therefore, they are functionalized or capped by gold (Au) or silver (Ag) in the form of core-shell structure. This is, as well, necessary for making the structure bio-compatible. In fact, attempts are being made \cite{Dabagh,Brollo,Felix,Anik} to cap or functionalize magnetic nanoparticles by Au or Ag for different biomedical applications. Such structures, however, more often than not exhibit decrease in magnetization because of the capping or functionalization \cite{Maenosono}. Motivated by these observations we attempted to examine whether capping by Ag could change the magnetization of nanoscale BiFeO$_3$ significantly and thus make them suitable for several applications including biomedical. This is all the more relevant for nanoscale BiFeO$_3$ since even bare nanoparticles of BiFeO$_3$ exhibit enhanced ferromagnetism at room temperature \cite{Park,Majumder,Carranza-Celis,Ramirez,Cardona-Rodriguez}. We report here that indeed decoration of multiferroic BiFeO$_3$ nanoparticles with Ag in the form of BiFeO$_3$/Ag nanocomposite results in an order of magnitude rise in saturation magnetization at room temperature than what is observed in bare BiFeO$_3$ nanoparticles. The charge transfer across the BiFeO$_3$/Ag interface and the specific pattern of self-assembly of the nanoparticles appear to be responsible for such an enormous enhancement in magnetization. This nanocomposite could be extremely useful for decoration of the malignant tumors necessary for efficacious hyperthermia. Because of magnetoelectric multiferroicity, the nanocomposite could be useful for ion channel gating action as well \cite{Caruntu} where remote application of magnetic field can locally trigger onset of finite electric potential if BiFeO$_3$/Ag based multiferroic ion channel gates replace the faulty ones.

\section{Experiments}
The BiFeO$_3$/Ag nanocomposite was prepared by taking Bi(NO$_3$)$_3$, 5H$_2$O and Fe(NO$_3$)$_3$, 9H$_2$O in 1:1 molar ratio and directly dissolving them in 8M potassium nitrate solution. When the solution becomes homogeneous with dark brown color, 4 vol\% of 5\% hydrogen peroxide H$_2$O$_2$ (30\% Merck) was added to the mixture and stirred well. The resultant solution was then transferred to a Teflon-lined autoclave for hydrothermal treatment for 30h at 180$^o$C. The powder was collected by centrifugation and washed several times with de-ionized water and ethanol. The as-prepared BiFeO$_3$ particles were dispersed in 1:4 de-ionized water and ethanol mixture and sonicated until they were dissolved; 20 wt\% silver nitrate (AgNO$_3$) was then added to the mixture. When the mixture became homogeneous, it was reduced by using NaBH$_4$ in 1:2 molar ratio. The solution immediately became dark brown and it was further stirred for 30 min. Finally, the powder was collected using a PVDF membrane filter of 0.22 $\mu$m pore size with the help of a vacuum filtration setup and washed by de-ionized water and ethanol mixture. 

The nanocomposite, thus prepared, was characterized by powder x-ray diffraction. The particle morphology, pattern of self-assembly, and the concentration distribution of the elements (Bi, Fe, O, Ag) across different regions of the nanocomposite were investigated by using transmission electron microscopy (TEM) and energy dispersive x-ray spectra (EDX). The surface/interface electronic structure of Bi, Fe, O, and Ag was determined by x-ray photoelectron spectroscopy (XPS). The magnetization was measured by a Vibrating Sample Magnetometer (LakeShore Model 7407). The magnetic domain structure was imaged at room temperature by magnetic force microscopy (LT AFM/MFM System of Nanomagnetics Instruments Ltd., Ankara, Turkey).  

\begin{figure}[ht!]
\centering
{\includegraphics[scale=0.35]{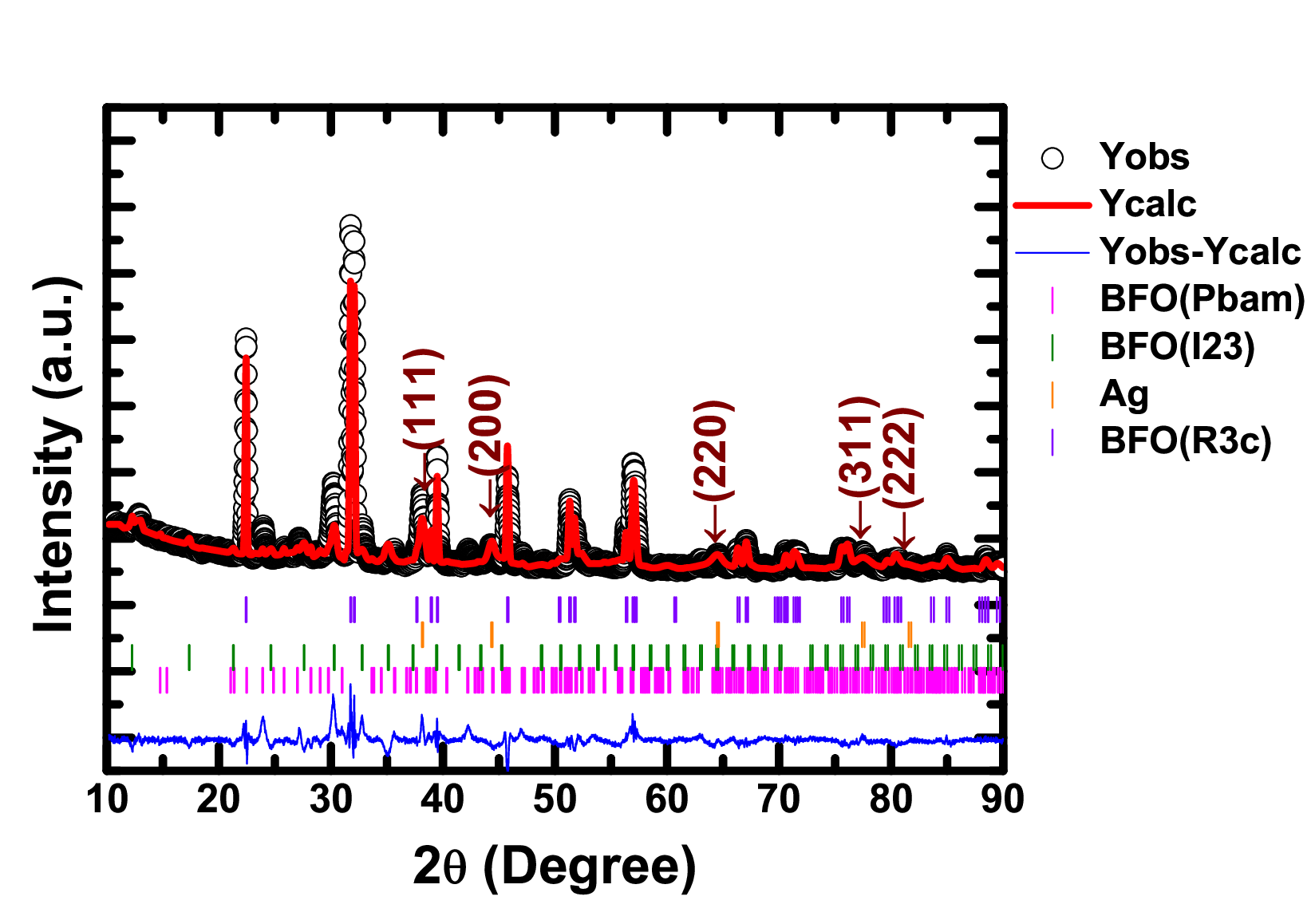}}
\caption{The room temperature powder x-ray diffraction (XRD) pattern and its refinement by FullProf. The peaks corresponding to Ag have been marked.}
\end{figure}

\begin{figure*}[ht!]
\centering
{\includegraphics[scale=0.60]{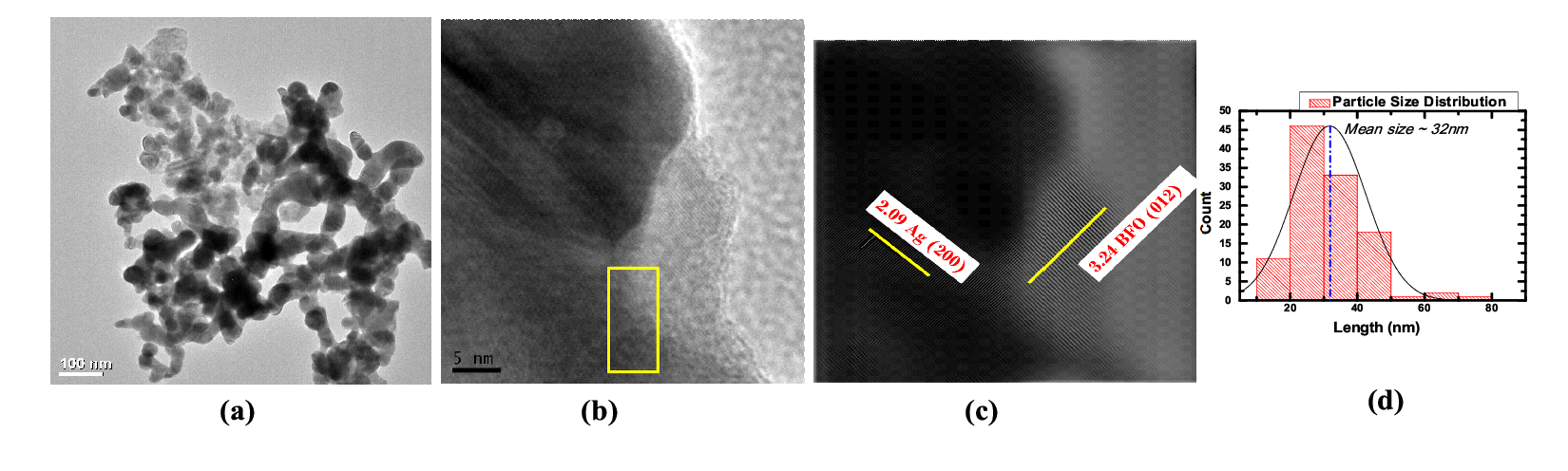}}
\caption{(a) Representative transmission electron microscopy (TEM) image of the BiFeO$_3$/Ag nanocomposite and (b), (c) high resolution TEM (HRTEM) image showing the presence of BiFeO$_3$ and Ag nanoparticles; image in (c) is obtained from the FFT and inverse FFT of the region shown in the box in (b); (d) the result of the TEM image analysis which shows that the average particle size is $\sim$32 nm. }
\end{figure*}

\section{Results and Discussion}
Figure 1 shows the room-temperature powder x-ray diffraction data of the BiFeO$_3$/Ag nanocomposite and their refinement by FullProf. The crystallographic structure turns out to be rhombohedral ($R3c$ space group) with lattice parameters $a$ = $b$ = 5.580 \AA and $c$ = 13.873 \AA. Peaks corresponding to Ag could also be observed. They are marked in Fig. 1. We used the JCPDS file 04-0783 for analyzing the peaks of Ag. Multiphase refinement yields the volume fraction of the respective phases - $\sim$65\% BiFeO$_3$ and $\sim$20\% Ag. Another sample with 2 vol\% Ag has also been prepared. The Ag assumes cubic structure (space group $Fm\bar{3}m$) with lattice parameter $a$ = 4.086 \AA. It is important to mention here that impurity Bi$_{25}$FeO$_{40}$ (space group $I23$)and Bi$_2$Fe$_4$O$_9$ ($<$5 vol\%) (space group $Pbam$) phases were found to be present in the nanocomposite. However, we prepared a reference BiFeO$_3$ nanoscale sample which also contains these impurity phases. The x-ray diffraction data and their refinement for the reference sample are included in the supplemental materials document. We compared the magnetization data of these two nanocomposites for highlighting the influence of Ag decoration. Comparison with the data of bare BiFeO$_3$ nanoparticles does not indicate formation of BiFeO$_3$/Ag solid solution. This is corroborated by the scanning transmission electron microscopy (STEM) and energy dispersive x-ray (EDX) mapping data (discussed later) where the concentration of Ag has been mapped across the nanoparticles. The high resolution TEM (HRTEM) data too point out that nanoparticles of Ag and BiFeO3 exists side-by-side; Ag does not affect the lattice of BiFeO$_3$. Interface interaction appears to be more important in the present case. The structural details such as the space group, lattice parameters, ion positions, as well as the fit statistics for both the samples (one which contains Ag and another which does not contain Ag) are available in the supplementary materials document (Table S-I).   

\begin{figure*}[ht!]
\centering
{\includegraphics[scale=0.50]{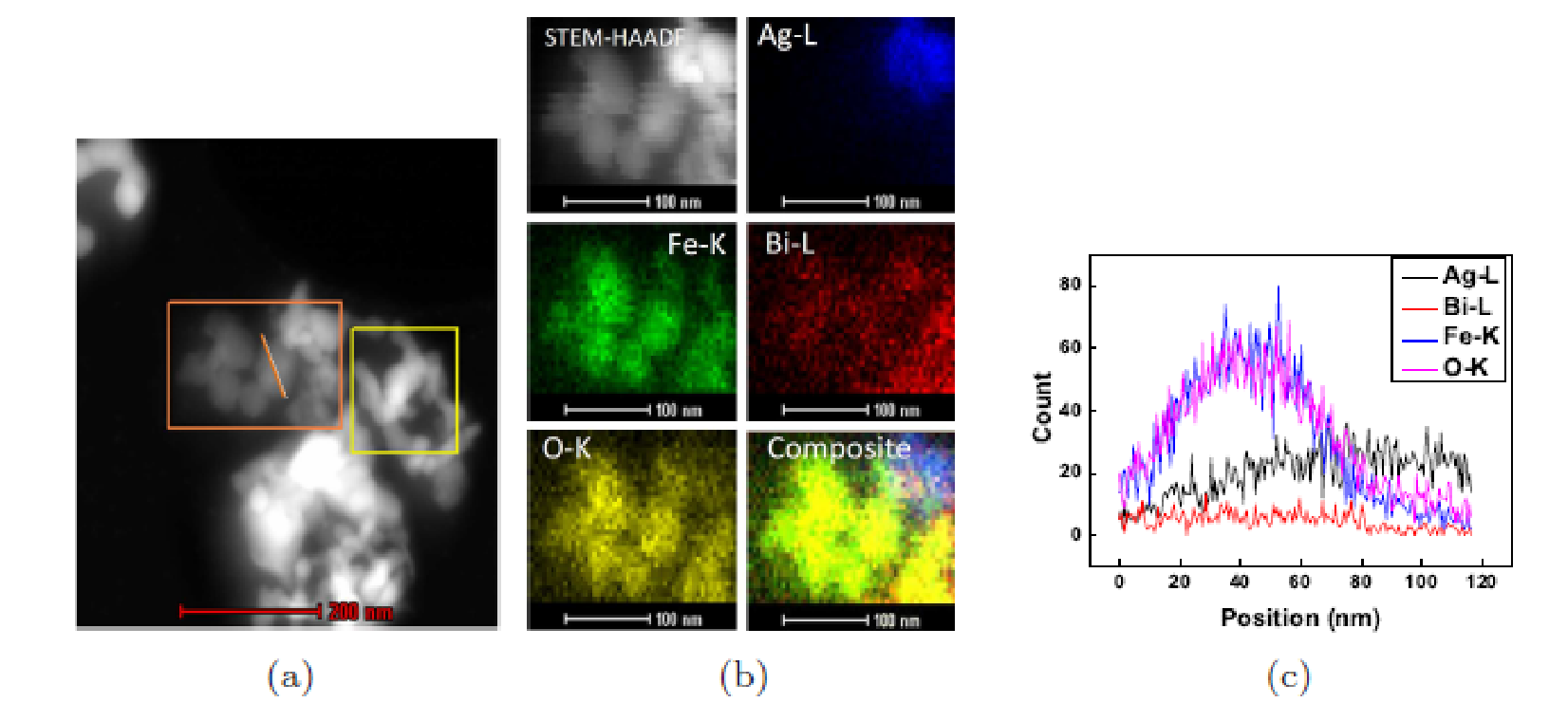}}
\caption{(a) Representative STEM-HAADF image of the BiFeO$_3$/Ag nanocomposite; (b) mapping of the distribution of element (Bi, Fe, O, and Ag) concentration across the nanocomposite obtained from the energy dispersive x-ray spectra recorded; (c) the mapping of the quantitative data on element concentration across the line shown in image (a).}
\end{figure*}

The particle morphology, self-assembly pattern, and the charge transfer across the BiFeO$_3$ and Ag nanoparticle interface were investigated in detail by TEM, EDX, and XPS. Figure 2 shows the representative bright-field TEM images of the nanocomposite across different length scales. The pattern of self-assembly and the particle size could be observed and determined from these images. The average size of the particles is $\sim$32 nm. It corroborates the result obtained ($\sim$35 nm) from the calculation of crystallite size using the line broadening of the x-ray diffraction data (supplementary materials document). It appears that a chain-like self-assembled structure has formed with finer Ag nanoparticles (average size $\sim$5 nm) surrounding the BiFeO$_3$ ones. The representative TEM image of the nanoscale BiFeO$_3$ reference sample (average particle size $\sim$32 nm) is included in the supplementary materials document. We recorded the high resolution TEM images across the BiFeO$_3$ and Ag nanoparticle interfaces. The images were fast Fourier transformed (FFT) and then inverted (inverse FFT) to generate clearer views of the lattice fringes. The corresponding `d' values were determined. It appears that (012) plane of BiFeO$_3$ and (200) plane of Ag nanoparticles are oriented perpendicular to the electron beam. Additional images showing the orientation of the (104) and (111) planes of, respectively, BiFeO$_3$ and Ag are included in the supplementary materials document. The scanning transmission electron microscopy (STEM) and the high-angle annular dark field (HAADF) images were also recorded at different regions of the nanocomposite (Fig. 3). The EDX mapping of the elements Bi, Fe, O, and Ag across those regions confirms the distribution pattern of the BiFeO$_3$ and Ag nanoparticles. In order to obtain a clearer view, the line scans of the element concentration were recorded across the interfaces between the BiFeO$_3$ and Ag nanoparticles at different regions of the nanocomposite (Fig. 3). Additional images and line scan data are available in the supplementary materials document. Combining the TEM, HRTEM, STEM images and the mapping of the concentration of the elements across the interfaces between BiFeO$_3$ and Ag nanoparticles by line scans it is possible to clearly notice how the nanoparticles are distributed or, in other words, the specific self-assembly patterns of the BiFeO$_3$ and Ag nanoparticles. Such a pattern of self-assembly appears to have maximized the interface area which, in turn, yields such enormous rise in saturation magnetization. 

\begin{figure*}[ht!]
\centering
{\includegraphics[scale=0.55]{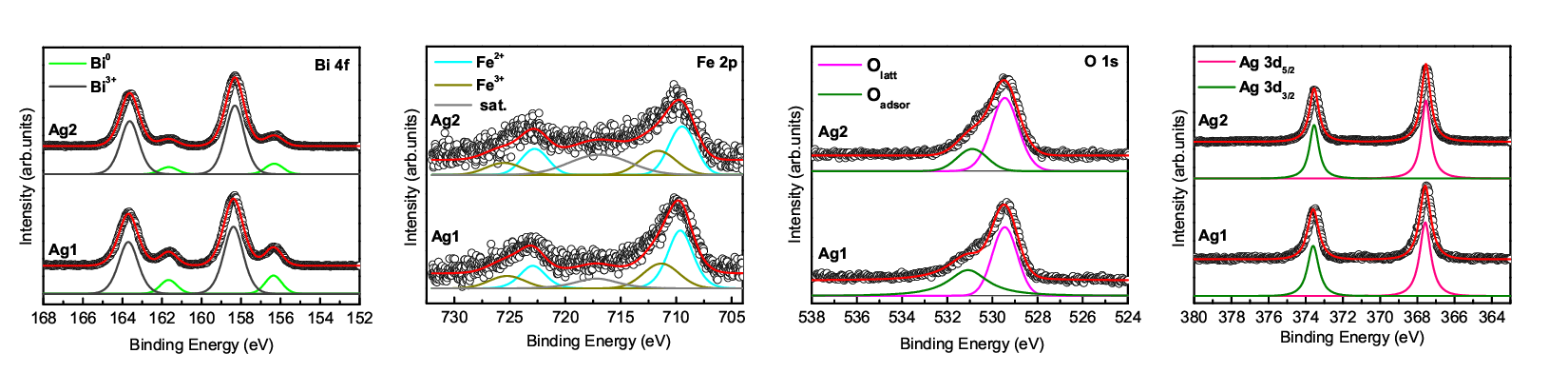}}
\caption{The x-ray photoelectron spectra for (a) Bi, (b) Fe, (c) O, and (d) Ag and their fitting.   }
\end{figure*}

\begin{figure}[ht!]
\centering
{\includegraphics[scale=0.25]{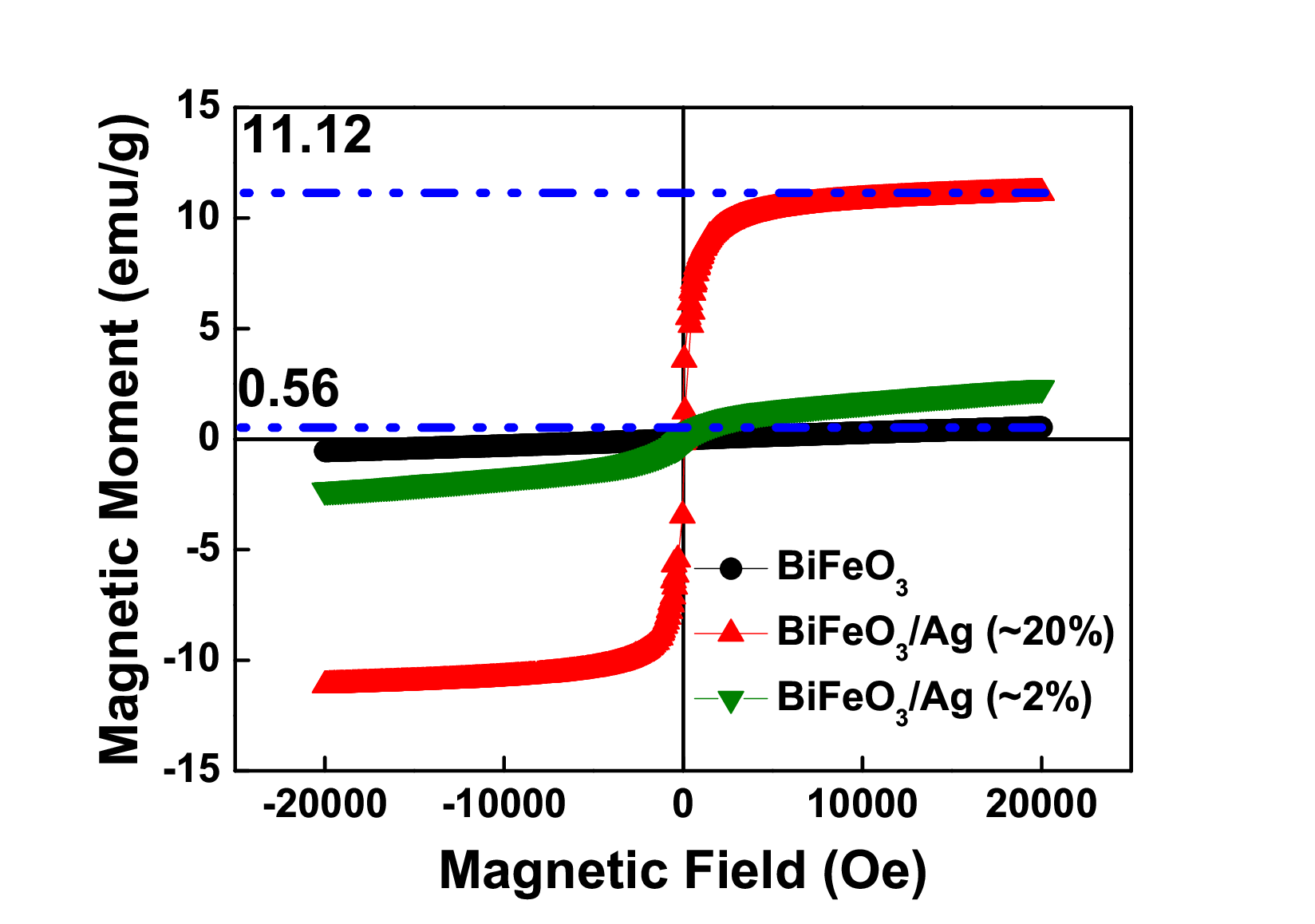}}
\caption{The room temperature $M$-$H$ hysteresis loops for the BiFeO$_3$ nanoparticles (reference sample) and the BiFeO$_3$/Ag nanocomposites containing $\sim$2 and $\sim$20 vol\% Ag.}
\end{figure}

The charge states of the ions have been examined by x-ray photoelectron spectroscopy (XPS). In Fig. 4, we show the spectra corresponding to the Bi, Fe, O, and Ag ions. The peaks have been fitted by using appropriate model following the background subtraction and tabulated in the Table-1. The peaks corresponding to Bi$^{3+}$ state are located at $\sim$158.3 and $\sim$163.7 eV, respectively. It indicates that no doping or substitution has taken place at the Bi-site. Small amount of metallic Bi$^0$ phase could be observed with corresponding peaks at $\sim$156.3 and $\sim$161.6 eV. The fitted Fe 2p spectra show the presence of characteristic doublet peaks of Fe(II)-O species at $\sim$709.5 and $\sim$723.0 eV for Fe 2p$_{3/2}$ and Fe 2p$_{1/2}$, respectively. The corresponding doublet peaks of Fe(III)-O species are present at $\sim$711.6 and $\sim$725.5 eV, respectively. The O 1s spectra contain characteristic features at $\sim$529.5 and $\sim$530.9 eV which correspond to the lattice oxygen and loss of oxygen, respectively. In the Ag 3d spectra, the characteristic doublet peaks could be observed at $\sim$367.5 and $\sim$373.6 eV. Comparison with the XPS spectra for pure BiFeO$_3$ and Ag shows the presence of a shift of $\sim$0.7 eV in the spectra of Ag and lattice oxygen of BiFeO$_3$. It points out orbital overlap and charge transfer between Ag and O states in the BiFeO$_3$/Ag nanocomposite. Contrary to the observations made in other cases, in the present case, electron transfer appears to have taken place from O to Ag creating holes in the O p states. The O 1s core level spectra of BiFeO$_3$/Ag nanocomposite clearly show an increase in high binding energy shoulder compared to that of BiFeO$_3$ sample, which is generally attributed to surface oxygen vacancies and/or adsorbed contaminant.

\begin{table}[ht]
{\caption {The XPS peaks for bare BiFeO$_3$, Ag, and BiFeO$_3$/Ag nanocomposite.}

\begin{tabular}{p{0.6in}p{0.7in}p{0.8in}p{0.6in}p{0.5in}} \hline\hline
Ion & Ionic state & BiFeO$_3$/Ag & Bare BiFeO$_3$ & shift  \\ \hline
Bi4f & Bi$^o$ & 156.3 & 156.8 & -0.5\\
     & Bi$^o$ & 161.6 & 162.1 & -0.5\\
     & Bi$^{3+}$ & 158.3 & 158.8 & -0.5\\
     & Bi$^{3+}$ & 163.7 & 164.1 & -0.5\\
Fe2p & Fe$^{2+}$ & 709.5 & 710.7 & -1.2\\
     & Fe$^{2+}$ & 723.0 & 724.2 & -1.2\\
     & Fe$^{3+}$ & 711.6 & 711.4 &  0.2\\
     & Fe$^{3+}$ & 725.5 & 725.3 &  0.2\\
     & Sat & 716.9 & 717.6 \newline 730.1\\
O1s  & Lattice O & 529.5 & 528.8 & 0.7\\
     & O vacancy & 530.9 & 529.7 & 1.2\\
     & Surface O & & 531.0 &\\
     &  & Ag & Bare Nano &\\
Ag3d & 3d 5/2 & 367.5 & 368.2 & -0.7\\
     & 3d 3/2 & 373.6 & 374.2 & -0.7\\
\hline \hline

\end{tabular}}
\end{table}

\begin{figure*}[ht!]
\centering
{\includegraphics[scale=0.50]{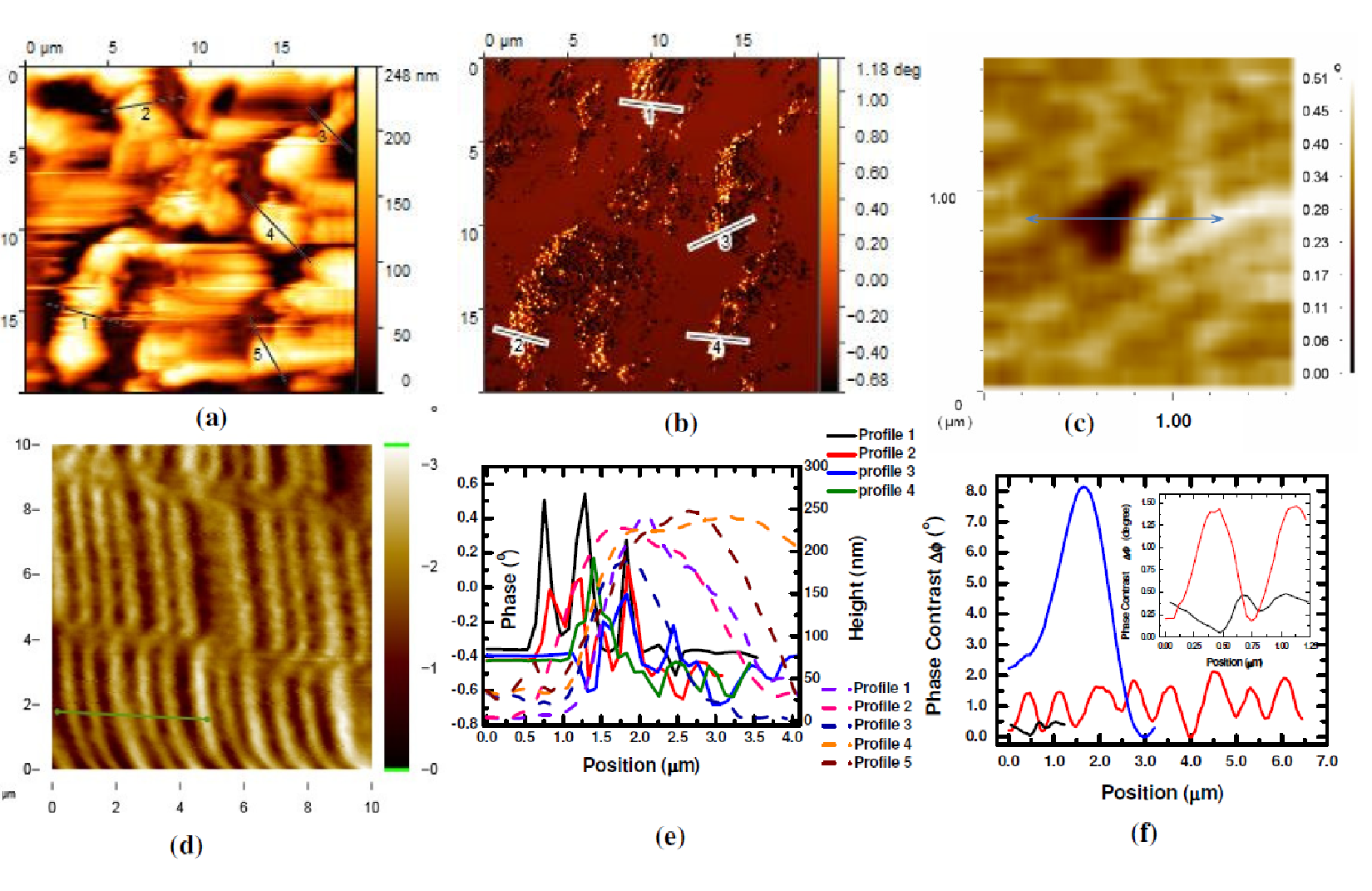}}
\caption{Representative magnetic force microscopy (MFM) (a) topography and (b) phase-contrast images of the BiFeO$_3$/Ag nanoclusters containg $\sim$20 vol\% Ag, (c) and (d) are the representative phase-contrast MFM images of the BiFeO$_3$ nanoparticle cluster and hard-disk surface commonly used in MFM as a standard sample; (e) line scan mapping of the topography and phase-contrast MFM images [corresponding lines are shown, respectively, in (a) and(b)] which indicates the size of the nanoparticle clusters and the magnetic domain size; the solid lines represent the magnetic domain size (left axis) while the dashed lines represent the size of the nanoparticle clusters (right axis); (f) line scan data with normalized phase contrast ($\Delta \phi$) for different samples - BiFeO$_3$/Ag nanocomposite (blue line), pure BiFeO$_3$ (black line), and the hard disk (red line); inset zooms in on the data for BiFeO$_3$ nanoparticles (black line) and hard disk (red line). }
\end{figure*}

Figure 5 shows the room temperature magnetization ($M$) versus applied magnetic field ($H$) magnetic hysteresis loops for the reference BiFeO$_3$ nanoparticles and BiFeO$_3$/Ag nanocomposites containing and $\sim$2 and $\sim$20 vol\% Ag. Clearly, more than an order of magnitude jump in saturation magnetization ($M_S$) could be observed [from $\sim$0.56 emu/g ($\sim$0.03 $\mu_B$/formula unit) to $\sim$11.12 emu/g ($\sim$0.6 $\mu_B$/formula unit)] in the nanocomposite with $\sim$20 vol\% Ag. The coercivity $H_C$, of course, turns out to be quite small ($\sim$60 Oe). Presence of Ag indeed gives rise to enhancement of ferromagnetism. We also measured the zero-field-cooled (ZFC) and field-cooled (FC) magnetization versus temperature patterns across 300-700 K. The result is included in the supplementary materials document. The transition temperature $T_N$ turns out to be $\sim$550 K. We further employed magnetic force microscopy (MFM) to observe the influence of large saturation magnetization on the phase contrast as well as on the magnetic domain structure. The nanocomposite (with $\sim$20 vol\% Ag) particles were dispersed in ethanol and deposited onto a glass substrate. The dried substrate was used for recording the MFM scan. Figure 6 shows the MFM topography and the phase-contrast images. They were processed by Gwyddion software. The line scans recorded across different regions of the phase-contrast image (Fig. 6b) reveals the size of the magnetic domains in different regions of the nanocomposite cluster. The lines are shown in the phase contrast image. The corresponding plot shows the average size of the domains (Fig. 6e). It turns out to be 1.0-1.5 $\mu$m. Therefore, across the particle clusters, the magnetic domains form by encompassing large number of individual particles. This has been observed by others as well \cite{Alivisatos}. We compare the phase-contrast data observed in this case with those obtained for the reference BiFeO$_3$ nanoparticle cluster and the standard magnetic thin film (surface of magnetic hard disk) (Fig. 6f) - black, red, and blue lines (Fig. 6f), respectively, represent the phase-contrast line profiles for the reference BiFeO$_3$ cluster, magnetic hard disk, and the BiFeO$_3$/Ag ($\sim$ 20 vol\%) nanocomposite cluster. The corresponding phase-contrast MFM images are shown in Figs. 6(c) and (d), respectively. Clearly, the phase contrast is enormous in the case of the BiFeO$_3$/Ag nanocomposite. Since, $\Delta\phi$ $\propto$ $K_p$.$\partial F_z/\partial z$ \cite{Mayergoyz,Asenjo}, large force gradient ($\partial F_z/\partial z$) resulting from large stray field gradient leads to large $\Delta\phi$. It is important to mention here that the same cantilever (Co-alloy coated Si cantilever-tip, coercivity $\sim$300 Oe) has been used to map the MFM images for all the samples - hard disk surface, BiFeO$_3$ nanoparticles, and BiFeO$_3$/Ag nanocomposite. Therefore, the constant factor $K_p$ remains invariant in all these cases. The plot in Fig. 6f shows that the $\Delta\phi$ in the nanocomposite is even higher than that in the hard disk. Therefore, the signature of large magnetization could be observed in MFM images as well. The crosstalk between topography and phase-contrast images in MFM, of course, often poses difficulty in imaging the magnetic domains clearly. However, since the magnetization is large in the present case (resulting in large stray field gradient and hence large phase contrast $\Delta\phi$) and dual pass technique with large cantilever lift height ($\sim$120-150 nm) has been used, the influence of topography on the phase-contrast images is minimal. The utility of dual pass tehnique and large cantilever lift height in eliminating the influence of topography on phase-contrast MFM images has been highlighted by others as well \cite{Freire}. The line scans, taken on the topography image (Fig. 6a), reveals the difference between the nanoparticle cluster size and the size of the magnetic domains (Fig. 6e). This difference highlights the negligible influence of topography on the phase-contrast MFM image. In fact, complete reversal of the magnetic domains could also be observed \cite{Goswami} even in bare BiFeO$_3$ nanoparticles upon reversal of applied magnetic field (+20 and -20 kOe). Additional MFM images of magnetic domain reversal due to reversal of applied magnetic field, observed in bare BiFeO$_3$ nanoparticles, are available in the supplementary materials document.

\begin{figure}[ht!]
\centering
{\includegraphics[scale=0.35]{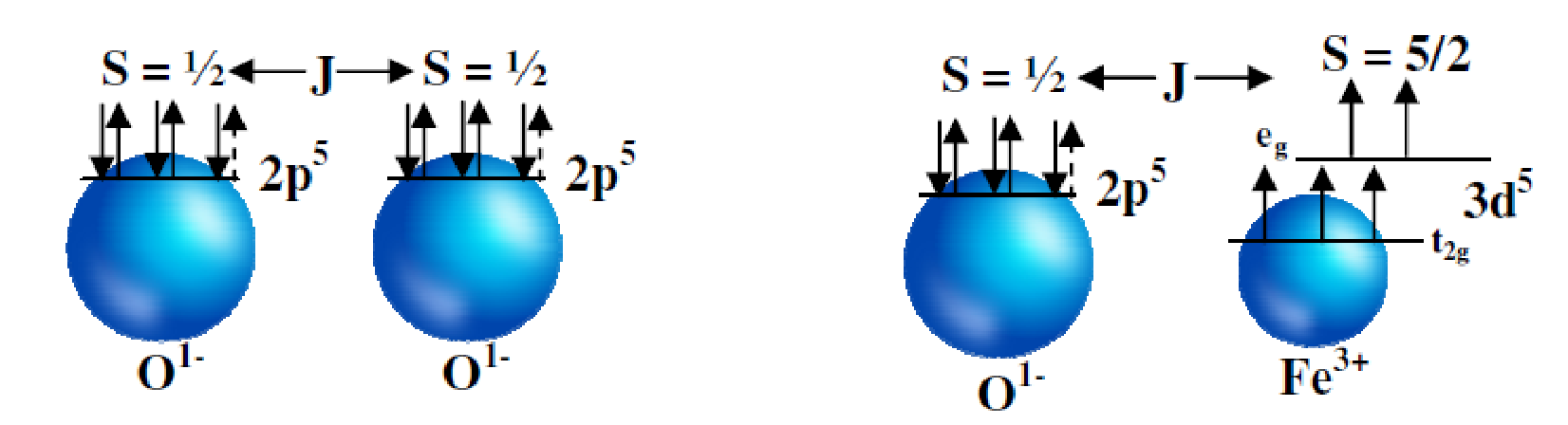}}
\caption{The schematic of the possible exchange coupling interactions across O$^-$-O$^-$ and O$^-$-Fe$^{3+}$ channels which indicate the additional magnetic interactions resulting from charge transfer between O and Ag.}
\end{figure}

Neither bare BiFeO$_3$ nanoparticles of average size $\sim$30 nm nor bare Ag$^0$ nanoparticles exhibit such a large magnetization. Though the ferromagnetism indeed depends on particle size in nanoscale BiFeO$_3$ \cite{Hasan}, we compared here magnetic properties of two samples of comparable average particle size ($\sim$30 nm) to rule out the influence of particle size. A cluster of Ag nanoparticles, on the contrary, is theoretically predicted \cite{Arias} to be giving rise to ferromagnetism because of charge transfer driven by symmetry of the cluster. Charge transfer between Ag 4d and O 2p states was also shown to be resulting in ferromagnetism if Ag films are exposed to oxygen \cite{Trudel}. In MoS$_2$/Ag nanocomposite, it was earlier observed \cite{Saha} that the holes in Ag 4d level and formation of Ag-S bonds (resulting from charge transfer between Ag and S) could induce ferromagnetism and metal-insulator transition. In the present case, the XPS data suggest that the charge transfer has taken place from O to Ag and thus creating holes in the O 2p states. As a result, it appears that alongside Ag 5s$^2$4d$^{10}$ states, O 2p$^5$ states ($S$ = $\frac{1}{2}$) could possibly emerge here in this nanocomposite from O 2p$^6$ states in pure BiFeO$_3$. The charge state shift, observed in Ag, has also been noticed in nanoscale Ag by others \cite{Huang}. In such a situation, in addition to the existing magnetic exchange interaction channels, exchange interactions could take place across the O$^-$-O$^-$ and Fe$^{3+}$-O$^-$ channels as well (Fig. 7). Therefore, p-state ferromagnetism seems to be assuming relevance in this case. Ferromagnetism due to p electrons in C around the vacancies in Si and C sites or due to trapping of holes in O sites nearest to the vacancies in Mg in MgO nanocrystals were earlier shown \cite{Wang,Choudhury} in SiC and MgO. Oxidization of oxygen and consequent trapping of holes or even molecular O$_2$ formation was earlier observed \cite{Bruce} in Na$_{0.6}$(Li$_{0.2}$Mn$_{0.8}$)O$_2$ system during charge-discharging cylces of a battery where Na$_{0.6}$(Li$_{0.2}$Mn$_{0.8}$)O$_2$ was used as the cathode. Of course, the oxidation states of Ag and O here and the origin of additional ferromagnetism, as a result, need to be probed in greater detail. This will be attempted in a future work. It is important to mention here that, in nanoscale magnetic oxides, the magnetization is generally expected to be influenced by uncompensated spins arising out of surface defects. The surface defects of Ag-free nanosized BiFeO$_3$ particles were earlier studied by XPS \cite{Chatterjee}. But, the magnetization data for the reference (i.e., Ag-free) BiFeO$_3$ sample clearly show that the BiFeO$_3$/Ag nanocomposite, comprising of nanoparticles of comparable size ($\sim$30-32 nm), exhibits more than an order of magnitude jump in magnetization. Therefore, the distinct influence of Ag and consequent charge transfer is discernible in the case of the nanocomposite. Further we point out that even though the Bi$_{25}$FeO$_{40}$ and Bi$_2$Fe$_4$O$_9$ phases are present in the nanocomposite, they do not contribute much to the room temperature ferromagnetism since the magnetic transition temperature ($T_N$) of these compounds are around 272 K \cite{Zahraouy} and 250 K, respectively. Moreover, these impurity phases are found to be present in the reference BiFeO$_3$ sample (which is used here for comparing the magnetization) as well.                  

\section{Summary}
In summary, we report observation of more than an order of magnitude jump in room-temperature saturation magnetization in BiFeO$_3$/Ag nanocomposite. This could be p-state ferromagnetism resulting from charge transfer between Ag and O. Both the saturation magnetization and the magnetic phase contrast (observed in MFM imaging) turn out to be quite enormous. Soft ferromagnetism with such a large saturation magnetization at room temperature could be extremely useful for various applications including biomedical, especially, decoration of malignant tumors necessary for efficacious treatment.\\

\noindent $\textbf{Acknowledgements}$\\

\noindent Two of the authors (TC, SM) acknowledge DST-INSPIRE fellowship, Govt of India, during this work. Another author (A.M.) acknowledges support fron Science and Engineering Research Board (SERB), Government of India (grant no. CRG/2021/002132).\\

\noindent $\textbf{Data availability statement}$\\

\noindent The data of this paper are available in the main paper as well as in the supplementary materials document (to be provided on request).\\   

\noindent $\textbf{References}$\\

\end{document}